\documentclass[runningheads]{llncs}

\usepackage{graphicx}
\usepackage{enumerate}
\usepackage{multirow}
\usepackage{array}
\usepackage{microtype}
\usepackage{subfigure}
\usepackage{cite}

\begin{document}
\title{SCGDet: Malware Detection using Semantic Features Based on Reachability Relation \thanks{Corresponding author: lurenjie17@mails.ucas.ac.cn}}
%
%
\author{Renjie Lu
} 
%
%
\authorrunning{Renjie Lu et al.}
%
\institute{University of Chinese Academy of Sciences, Beijing, China\\
\email{lurenjie17@mails.ucas.ac.cn}
}
\maketitle              
\begin{abstract}
Recently, with the booming development of software industry, more and more malware variants are designed to perform malicious behaviors. The evolution of malware makes it difficult to detect using traditional signature-based methods. Moreover, malware detection has important effect on system security. In this paper, we present SCGDet, which is a novel malware detection method based on system call graph model (SCGM). We first develop a system call pruning method, which can exclude system calls that have little impact on malware detection. Then we propose the SCGM, which can capture the semantic features of run-time program by grouping the system calls based on the reachability relation. We aim to obtain the generic representation of malicious behaviors with similar system call patterns. We evaluate the performance of SCGDet using different machine learning algorithms on the dataset including 854 malware samples and 740 benign samples. Compared with the traditional n-gram method, the SCGDet has the smaller feature space, the higher detection accuracy and the lower false positives. Experimental results show that SCGDet can reduce the average FPR of 14.75\% and improve the average Accuracy of 8.887\%, and can obtain a TPR of 97.44\%, an FPR of 1.96\% and an Accuracy of 97.78\% in the best case.

\end{abstract}
\section{Introduction}
Malicious software is referred to as malware, which is designed to perform various malicious activities, such as leaking private information, disabling targeted host and so on. Nowadays, the exponential growth of malware is a major threat in the software industry. McAfee, an anti-malware vendor \cite{r1}, reported that the total number of malware samples has grown almost 34\% over the past quarters to more than 774 million samples. Meanwhile, with the booming development of software industry, the malware has greatly evolved and become very sophisticated.

Given the alarming growth of malware, a large number of researches have focused on proposing malware detection techniques. Roughly, the malware detection techniques can be divided into two broad categories: static malware analysis and dynamic malware analysis. Traditional anti-virus software usually depends on static signature-based method to detect the malware \cite{r2}. Though these static methods can detect malicious applications before they are executed. However, the static malware analysis methods can be evaded using the encryption, the obfuscation or the packing \cite{r3}. Malware authors can write several malware variants that have similar functionality but different signatures to evade static malware detection methods, and the zero-day malware can also evade static malware analysis.

In response to this, various dynamic malware detection approaches have been proposed \cite{r4, r5, r6, r7, r8}, which focus on the program behaviors during execution. In particular, these techniques inspect the run-time behaviors of program by analyzing the system calls. The principal dynamic behavior-based malware detection methods include function call monitoring \cite{r4}, information flow tracking \cite{r5}, sequence modeling of system calls using n-gram model \cite{r6}, individual system call analysis \cite{r7} and behavioral graphs \cite{r8}. Though these approaches can detect the obfuscated malware and the variants of malware, they require a large amount of time and resources to carry out the detection. Moreover, the number of features generated by these approaches is huge, which results in the scalability of these approaches is highly problematic..

In this paper, we propose the SCGDet that exploits the following two observations to perform malware detection. \textit{Experimental observation 1}: The malware usually invokes security-critical system calls to achieve malicious activities. For example, the Brk exploit uses repetitively brk system call to increase the data segment of user program and overloads the main memory. \textit{Experimental observation 2}: Malware and its variants belonging to the same family usually invoke similar system call patterns due to performing the similar malicious behaviors. Based on the above observations, we propose the system call graph model (SCGM) to capture the semantic features of malicious behaviors. Our work aims to obtain the generic feature representation of malicious behaviors with similar system call patterns. Therefore, our approach has the potential to detect malware variants. We apply different machine learning algorithms to perform malware detection, and evaluate the performance of SCGDet using the dataset including 854 malware samples and 740 benign programs.
In summary, we make the following contributions in this paper:
\begin{enumerate}
\item We propose SCGDet, which is a novel malware detection method based on SCGM. The SCGM can capture the semantic features of programs by grouping the system calls based on the reachability relation and considering the frequency of these security-critical system call patterns.
\item Based on statistical experiments, we propose and formalize a novel system call pruning method in order to exclude system calls that have few effect on malware detection. 
\item We evaluate the performance of SCGDet using benign and malware samples. Experimental results show that SCGDet can obtain a TPR of 97.44\%, an FPR of 1.96\% and an Accuracy of 97.78\% in the best case. Compared with the traditional n-gram technique, experimental results demonstrate that our approach reduces significantly the feature space, have the lower false positives and the higher detection accuracy. Specifically, SCGDet can reduce the average FPR of 14.75\% and improve the average Accuracy of 8.887\%.
\end{enumerate}

The rest of this paper is organized as follows. Section II introduces our proposed malware detection method in detail. Experiment and evaluation are presented in Section III, and Related works are discussed in Section IV. Section V concludes the paper and future work.
\section{Methodology of SCGDet}
In this section, we introduce the proposed malware detection technique in detail. Figure 1 shows the overall architecture of SCGDet, which consists of training stage and testing stage. The training stage contains the system call tracing and pruning, the feature construction and generation and the malware detection model generation. Then, we use this malware detection model to predict whether the program under detection is malware during the testing stage.
\begin{figure}
\includegraphics[width=\textwidth]{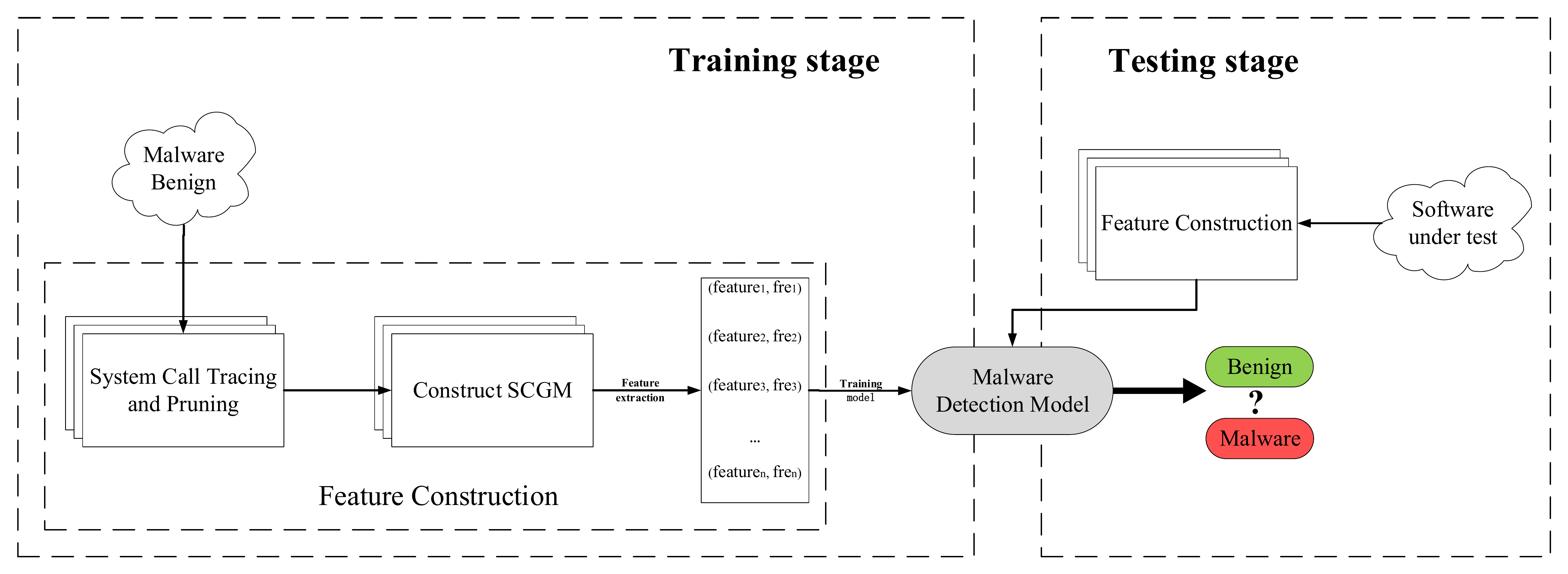}
\caption{The overview of SCGDet.} \label{fig1}
\end{figure}
\subsection{System Call Tracing and Pruning}
\subsubsection{System call tracing}
To generate system call trace records, all malware samples were executed in a controlled and virtualized environment running 64-bit Ubuntu 18.04 with Linux 4.15 Kernel. We use the \textit{strace}, which is a Linux debugger to trace system calls and signals, to generate system call trace records of each malware sample. In contrast, all benign samples were executed in a normal environment and common operations were performed on the benign programs. In a similar way, we also use strace to generate system call trace records of each benign sample.

\subsubsection{System call pruning}
Generally, no program requests all the system calls. Moreover, when we analyze a large number of programs, the total number of system calls requested by all the programs will be huge, which can result in the considerable time overhead. Therefore, we first identify the valuable system calls to eliminate the need of considering all system calls. We propose the following two rules to filter out system calls that have little impact on malware detection.

\textit{Rule 1: Exclude system calls that are commonly requested by both malware and benign samples} Each system call describes a particular operation, and the benign programs and malicious programs should request different system calls corresponding to their operational needs. That is, we do not need to analyze all system calls to bulid the malware detection model. As a result, we should exclude system calls that are commonly used by both benign and malicious programs, For example, system call \textbf{close} are frequently requested by both malicious and benign programs.

For this purpose, we first create two matrices and a list, M, B and SC. SC contains all the different system calls that appear in the dataset. M is a matrix of system calls used by malware samples, and B is a matrix of system calls used by benign programs. More formally, SC, M and B are shown as follows, respectively, where l represents the number of different system calls, m (854) indicates the number of malware samples in the dataset and n (740) indicates the number of benign samples in the dataset. $M_{ij}$ represents whether the j-th system call is requested by the i-th malware sample, and $B_{ij}$ represents whether the j-th system call is requested by the i-th benign sample, while '1' indicates yes and '0' indicates no.
\begin{equation}
SC = 
\left[
\begin{array}{cccccc}
 SC_1 & SC_2 & \cdots & SC_i & \cdots & SC_l\\
\end{array}
\right]
\end{equation}

\begin{equation}
M = 
\left[ 
\begin{array}{cccc}
 M_{11} & M_{12} & \cdots & M_{1l} \\
 M_{21} & M_{22} & \cdots & M_{2l} \\
 \vdots & \vdots & \ddots & \vdots \\
 M_{m1} & M_{m2} & \cdots & M_{ml} \\
\end{array}
\right]
\end{equation}

\begin{equation}
B = 
\left[ 
\begin{array}{cccc}
 B_{11} & B_{12} & \cdots & B_{1l} \\
 B_{21} & B_{22} & \cdots & B_{2l} \\
 \vdots & \vdots & \ddots & \vdots \\
 B_{n1} & B_{n2} & \cdots & B_{nl} \\
\end{array}
\right]
\end{equation}

Then we use the following formula to calculate the distribution of each system call, where $\frac{m}{n}$ is used to balance the effect of different sizes of malware dataset and benign dataset, and $D(SC_j)$ represents the distribution of the j-th system call. The result of $D(SC_j)$ ranges from 1 to -1. If $D(SC_j)$ = 1, this means that system cal $SC_j$ is only requested by the malware samples. If $D(SC_j)$ is close to 0, this means that $SC_j$ have little impact on malware detection because $\sum_{i=1}^m M_{ij}$ is closely equal to $\sum_{i=1}^n B_{ij}$. If $D(SC_j)$ = -1, this means that system call $SC_j$ is only requested by the benign samples. Finally, we filter out those system calls that the value of distribution is closely equal to 0.
\begin{equation}
    D(SC_j) = \frac{\sum_{i=1}^m M_{ij} - \sum_{i=1}^n B_{ij} * \frac{m}{n}}{\sum_{i=1}^m M_{ij} + \sum_{i=1}^n B_{ij} * \frac{m}{n}}
\end{equation}

\textit{Rule 2:Exclude system calls that appear only in few samples} To further filter out the system calls that are not helpful for malware detection. We do not consider those system calls to build malware detection model, which appear only in few samples. We use the following formula to measure this rule, where $N(SC_j)$ indicates the number of malware and benign samples using the j-th system call. The result of $N(SC_j)$ has a value ranging between 0 and $m+n$. Similarly, we filter out the system calls that the value of $N(SC_j)$ is closely equal to 0. For example, the system call \textbf{signal} only requested by one sample, so we should filter out it according to \textit{Rule 2}.
\begin{equation}
    N(SC_j) = {\sum_{i=1}^m M_{ij} + \sum_{i=1}^n B_{ij}}
\end{equation}

Finally, we neglect those system calls that have little impact on malware detection and consider only these meaningful and security-critical system calls, as listed in Table 1. It's worth noting that a file is an instance of any operated file or I/O device. We observe that the combination of these system calls (in other words, system call pattern) can result in the specific malicious activities. For example, the virus first copies its content into the temporary file, then changes the permission and the accessed time, and finally executes the modified file. Therefore, we can capture specific malicious behaviors of malware samples by analyzing the system call patterns.

\begin{table}
\centering
\caption{Common security-critical system calls in our dataset}\label{tab1}
\begin{tabular}{|p{2.8cm}<{\centering}|p{9cm}<{\centering}|}
\hline
Resource Types & Related system calls\\
\hline
File System &  open, openat, read, write, close, select, stat, fstat, lstat, statfs, stat64, fstat64, readlink, access, fcntl, ioctl, chdir, fchdir, getdents64, getcwd, lseek, utime, uname, unlink, umask, chmod, rename, execve\\
\hline
Process &  rt\_sigaction, rt\_sigprocmask, fork, rt\_sigsuspend, getpid, clone, waitpid, nanosleep, set\_tid\_address, prctl, getppid, pipe, kill\\
\hline
Network &  socket, connect, setsocket, bind, getsockname, listen \\
\hline
Memory &  brk, mmap, mmap2, munmap, mprotect\\
\hline
\end{tabular}
\end{table}

\subsection{Feature Construction: System Call Graph Model}
Feature construction is a crucial step of our proposed malware detection approach. The performance of any malware detection models depends heavily on how precisely the features represent the characteristics of the samples. In this work, we propose a novel feature construction method, system call graph model (SCGM), to capture the semantic features of typical malware behavior. We will detailedly illustrate the SCGM with the system call trace shown in Figure 2.
\begin{figure}
\centering
\includegraphics[width=11cm, height=7.5cm]{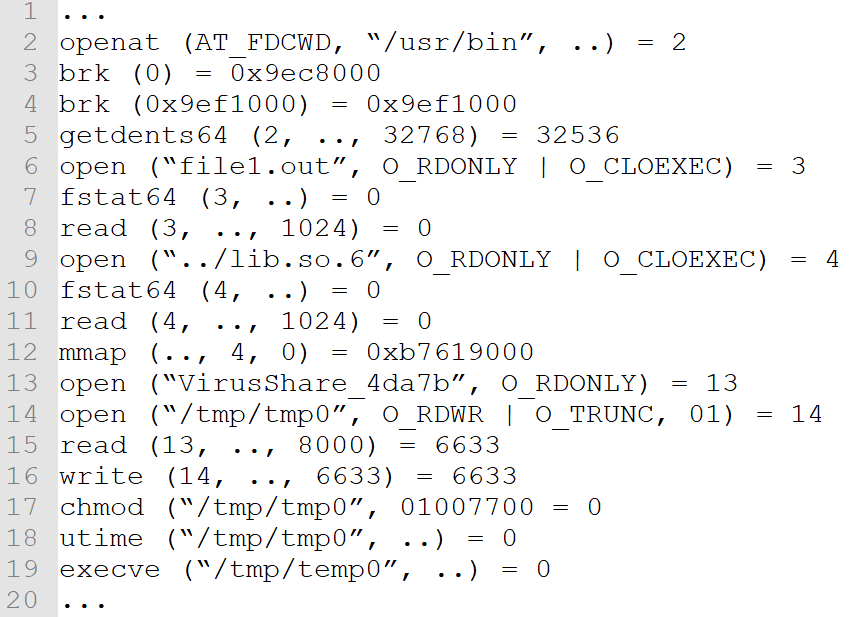}
\caption{A system call trace example.} \label{fig1}
\end{figure}

The SCGM is based on the high-level semantic features of malicious behaviors. We use security-critical system calls along with their arguments and return values to analyze malicious behaviors. We aim at to obtain the generic representation of malicious behaviors in order to detect the variants of the malware that have similar system call patterns. Meanwhile, we also consider the frequency of security-critical system call patterns for feature construction in order to capture malicious behaviors that execute repetitively specific system call pattern to overload the system. In short, we propose the following properties to capture the high-level semantic features of malicious behaviors.

\textit{Property 1: Group system calls based on reachability relation}
The observation behind this property is that malware and its variants usually perform a series of similar operations on particular resources in order to accomplish similar malicious behaviors. For example, in the system call trace shown in Figure 2, the virus first copies its content into the temp0 file, then changes the permission and the accessed time, and finally executes the modified file. These specific system call patterns can reveal malicious intents of malware. To group system calls based on reachability relation, we redeclare the concept of system call graph first.

In our work, the system call graph (SCG) can represent all relationships among system calls during program execution. The SCG is a directed graph and SCG = (V, Entity, E), where V is the set of various system calls, Entity is the set of source and destination files, and E is the set of edges. More formally, E = $\left\{<n_1,n_2> | n_1 \in (V \cup Entity), n_2 \in V\right\}$. That is, the edge $<n_1, n_2>$ is an out-edge of vertex or entity $n_1$ and in-edge of vertex $n_2$. 

In the system call trace example shown in Figure 2, V = $\left\{openat, brk, ..., utime,\right.$
$\left.execve\right\}$, Entity = $\left\{`/usr/bin', `file1.out', ..., `VirusShare\_4da7b', `/tmp/temp0'\right\}$, 
and E = $\left\{<`/usr/bin', openat>, <openat, getdents64>, <`file1.out', open>,\right.$
$\left. ..., <`/tmp/temp0', utime>, <`/tmp/temp0', execve>\right\}$ as shown in Figure 3.

Based on the SCG, we consider that the vertices that satisfy the reachability relations shown in Figure 4 should be grouped in the same set. \textit{1)} if $v_1$ and $v_2$ are directedly connected in SCG, $v_1$ and $v_2$ should be grouped in same set. \textit{2)} if $v_1$ and $v_2$ are directedly connected and $v_1$ and $v_3$ are also directedly connected in SCG, $v_1$, $v_2$ and $v_3$ should be grouped in same set. \textit{3)} if $v_1$ and $v_3$ are directedly connected and $v_2$ and $v_3$ are also directedly connected in SCG, $v_1$, $v_2$ and $v_3$ should be grouped in same set. \textit{4)} if $entity_1$ and $v_1$ are directedly connected and $entity_1$ and $v_2$ are also directedly connected in SCG, $v_1$ and $v_2$ should be grouped in same set.

\textit{Property 2: Counting the frequency of security-critical system call patterns}
The observation behind this property is that malware often executes repetitively security-critical system call patterns in order to overload the system resources (such as Memory, CPU), which can result in the denial of service, the CPU-starvation and even the crashing of OS in resource limited systems. For example, the \textit{Brk.c} exploit invokes repetitively \textit{brk} system call to increase the data segment of user program in order to overloads the main memory. Therefore, recording the frequency of these security-critical system call patterns is important to identify similar malicious behaviors.
\begin{figure}
\centering
\includegraphics[width=12cm, height=6.5cm]{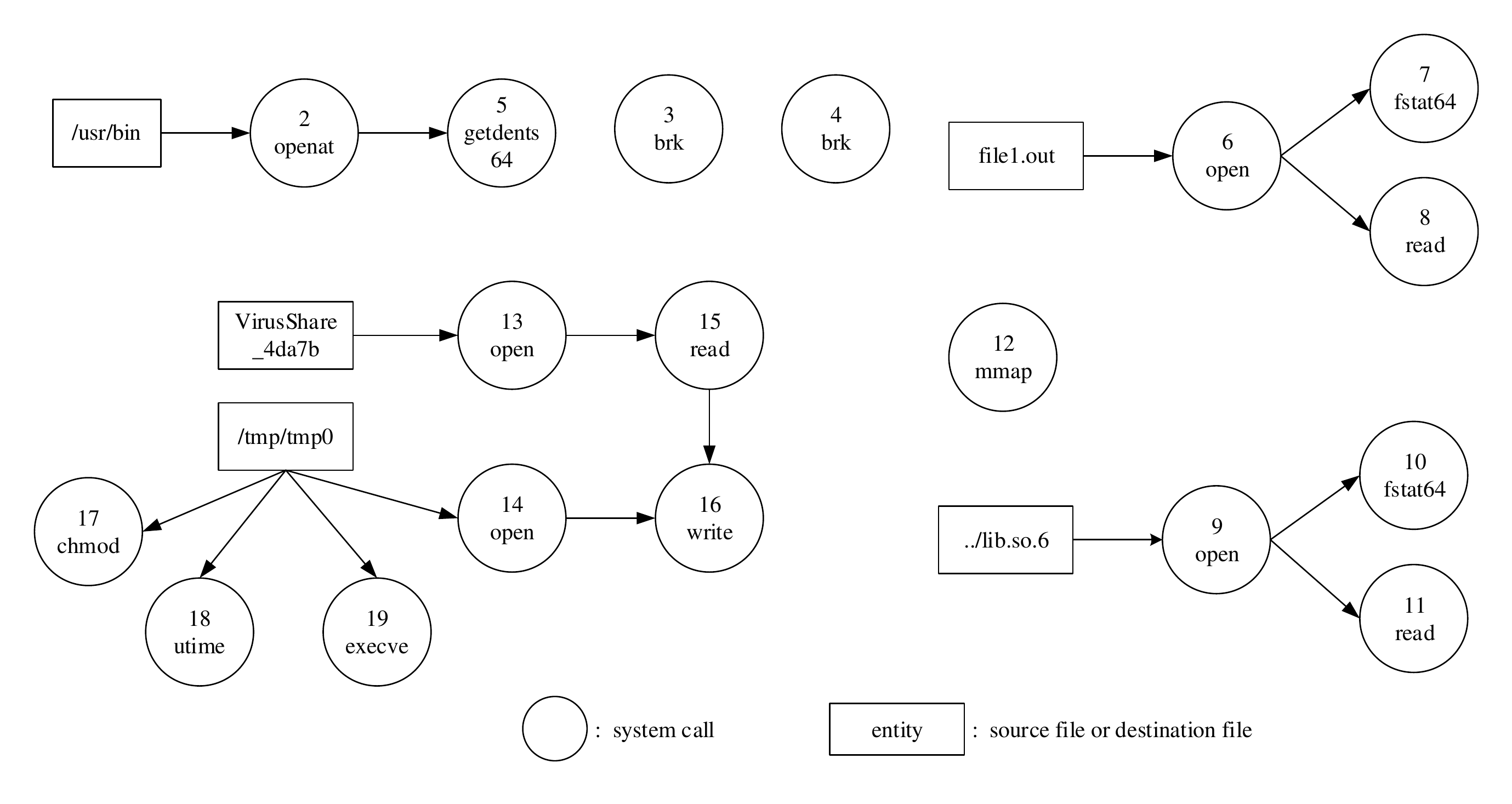}
\caption{The SCG corresponding to system call trace example shown in Figure 2.} \label{fig1}
\end{figure}
\begin{figure}
\includegraphics[width=\textwidth]{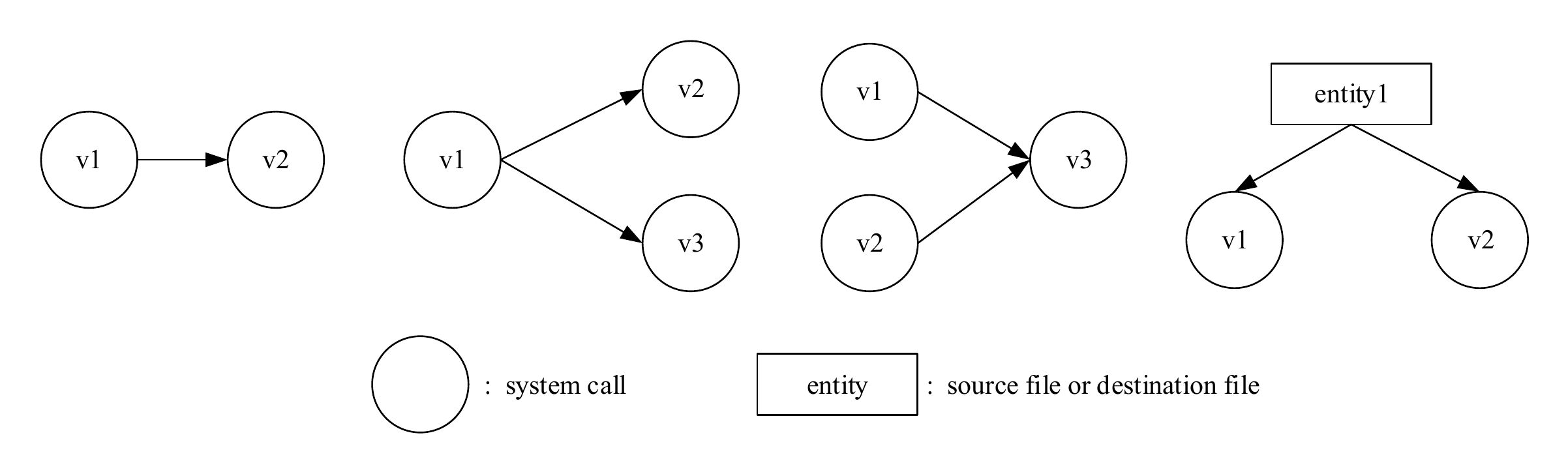}
\caption{Four reachability relations (structures) in the SCG.} \label{fig1}
\end{figure}

Hence, for the system call trace example shown in Figure 2, the feature set generated by our proposed SCGM contains $\left\{brk\right\}$, $\left\{mmap\right\}$, $\left\{open, fstat64, read\right\}$, $\left\{openat, getdents64\right\}$ and $\left\{open, open, read, write, chmod, utime, execve\right\}$, where the frequency of system call patterns $\left\{brk\right\}$ and $\left\{open, fstat64, read\right\}$ are equal to 2.

\subsection{Feature Vector Generation}
We utilize these system call patterns generated by the SCGM to construct a feature vector of size N for each sample, where N is the total number of different system call patterns (features) obtained from both malware samples and benign samples. As listed in Table 2, the feature vector FV = $\left\{<Feature_1, f_1>, <Feature_2, f_2>,\right.$
$\left. ..., <Feature_N, f_N>,\right.$
$\left.<Class, 1/0>\right\}$, where f indicates the frequency of this unique system call pattern (feature) in the current sample. If f is equal to 0, which indicates the corresponding system call pattern does not appear in the sample. It is worth noting that the last feature of FV is the label of sample, and 1 for malware and 0 for benign.

\begin{table}
\centering
\caption{The example of the feature vector}\label{tab1}
\begin{tabular}{|p{2cm}<{\centering}|p{1.5cm}<{\centering}|p{1.5cm}<{\centering}|p{1.5cm}<{\centering}|p{1.5cm}<{\centering}|p{1.5cm}<{\centering}|p{1.5cm}<{\centering}|}
\hline
Feature & $Feature_1$ & $Feature_2$ & $Feature_3$ & ... & $Feature_N$ & Class\\
\hline
Frequency & $f_1$ & $f_2$ & $f_3$ & ... & $f_N$ & 1/0 \\
\hline
\end{tabular}
\end{table}

\section{Experiments and Evaluations}
In this section, we first introduce the dataset and the evaluation metrics used in our experiments. Then, we present the experimental results in detail. We conduct lots of experiments to evaluate the effectiveness of SCGDet in distinguishing between malware and benign programs. We use python language to implement the system call tracing and pruning as well as feature construction, and use scikit-learn toolkit \cite{r9} to obtain malware detection model.

\subsection{Dataset and Metrics}
We collect 854 malware samples from VirusShare \cite{r10}, which include trojan, exploit, virus and various malicious scripts. We use the Virustotal \cite{r11} online tool to identify the types of these malware samples. In addition, we also collect 740 benign programs on Ubuntu 18.04 OS, which contains internet, system tools, video, office and so on. We divided the entire dataset into 70\%-30\%. 70\% of the samples were used to generate the malware detection model using the standard 10-fold cross-validation approach. The remaining 30\% of the samples were used to evaluate the performance of malware detection model.

For the malware detection, we first introduce the confusion matrix as shown in Table 3. Finally, we use the following evaluation metrics in our experiments: $True$ $Positive$ $Rate$ $(TPR)$ = $\frac{TP}{TP + FN}$, $False$ $Positive$ $Rate$ $(FPR)$ = $\frac{FP}{FP + TN}$ and $Accuracy$ $(ACC)$ = $\frac{TP + TN}{TP + FN + FP + TN}$.

\begin{table}[htbp]
\caption{The confusion matrix for malware detection}
\begin{center}
\begin{tabular}{|p{2cm}<{\centering}|p{1.5cm}<{\centering}|p{3cm}<{\centering}|p{3cm}<{\centering}|}
\hline
\multicolumn{2}{|c|}{\multirow{2}*{}} & \multicolumn{2}{|c|}{Predicted class} \\
\cline{3-4}
\multicolumn{2}{|c|}{~} & Malware & Benign \\
\hline
\multirow{2}*{Actual class} & Malware & True Positive (TP) & False Negative (FN) \\
\cline{2-4}
~ & Benign & False Positive (FP) & True Negative (TN) \\
\hline
\end{tabular}
\end{center}
\end{table}

\subsection{Experimental results}
We apply three widely used machine learning algorithms to build malware detection models: Logistic Regression (LR), Support Vector Machine (SVM) and Random Forest (RF). In order to compare the malware detection performance with other methods, we also implement 4-gram model on our dataset, which has better performance as presented in \cite{r6} and \cite{r12}.

\subsubsection{Feature Space Analysis}
Firstly, we analyze the difference in the feature space obtained by the 4-gram method and our proposed feature construction method, SCGM. For this purpose, we randomly select 25\%, 50\% and 75\% of the samples in our dataset for feature construction using 4-gram modeling technique and SCGM, respectively. Then we repeatedly perform this process five times and select all samples (just one time) in our dataset for feature construction. we record the sizes of different feature spaces and the average sizes of different feature spaces, as shown in Table 4 and Table 5. We visualize the difference in the average size of feature space between 4-gram model and SCGM, as shown in Figure 5. In can be seen that, for 4-gram model, the number of features grows proportionally with the number of samples. However, for SCGM, we observe that the curve gradually flattens as the number of samples increases.
\begin{table}
\centering
\scriptsize
\caption{The size of different feature spaces generated by different numbers of samples}
\begin{tabular}{|c|c|c|c|c|c|c|c|c|c|c|c|c|c|c|c|c|}
\hline
 & \multicolumn{16}{|c|}{The size of feature space}\\
\hline
Samples & \multicolumn{5}{|c|}{25\%} & \multicolumn{5}{|c|}{50\%} & \multicolumn{5}{|c|}{75\%} & 100\% \\
\hline
4-gram	& 4935	& 5940	&5195	&4770 & 5500	&6195	&5900	&6040	&6015	&6405	&6995	&6895	&6515	&6560	&6890	&7305 \\
\hline
SCGM & 150	&152 &150	&169	&153	&178	&171	&179	&174	&178	&181	&189	&190	&184	&186	&191\\
\hline
\end{tabular}
\end{table}

\begin{table}
\centering
\caption{The average size of different feature spaces listed in Table 4}
\begin{tabular}{|p{2cm}<{\centering}|p{1.5cm}<{\centering}|p{1.5cm}<{\centering}|p{1.5cm}<{\centering}|p{1.5cm}<{\centering}|}
\hline
 & \multicolumn{4}{|c|}{The average sizes of feature space}\\
\hline
Samples & 25\% & 50\% & 75\% & 100\% \\
\hline
4-gram	& 5268	& 6111	& 6771	& 7305 \\
\hline
SCGM & 155	& 176	& 186	& 191\\
\hline
\end{tabular}
\end{table}

\begin{figure} 
\subfigure[4-gram] {
 \label{fig:a}     
\includegraphics[width=0.5\columnwidth, height=4.7cm]{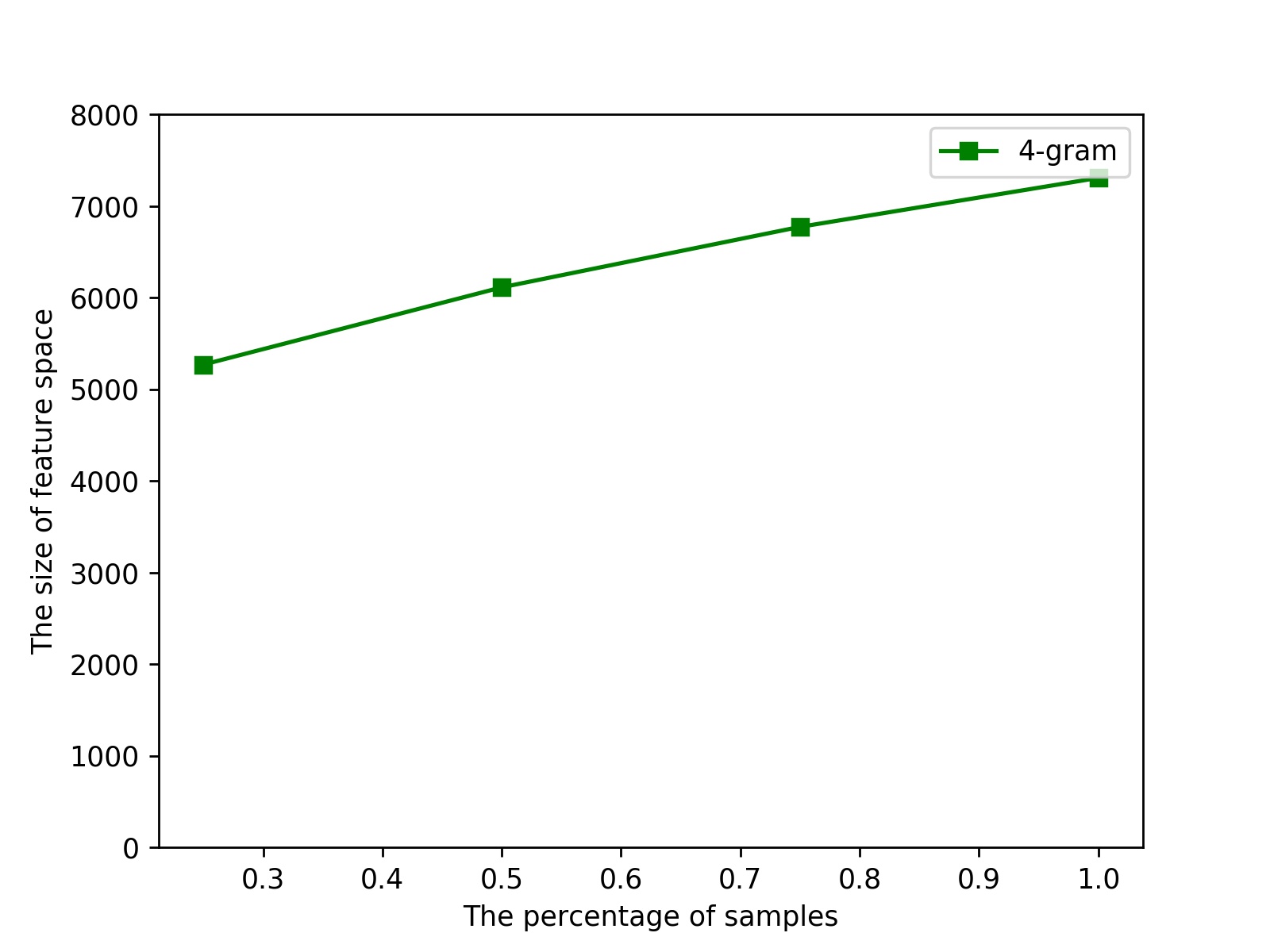}}  
\subfigure[SCGM] { 
\label{fig:b}     
\includegraphics[width=0.5\columnwidth, height=4.7cm]{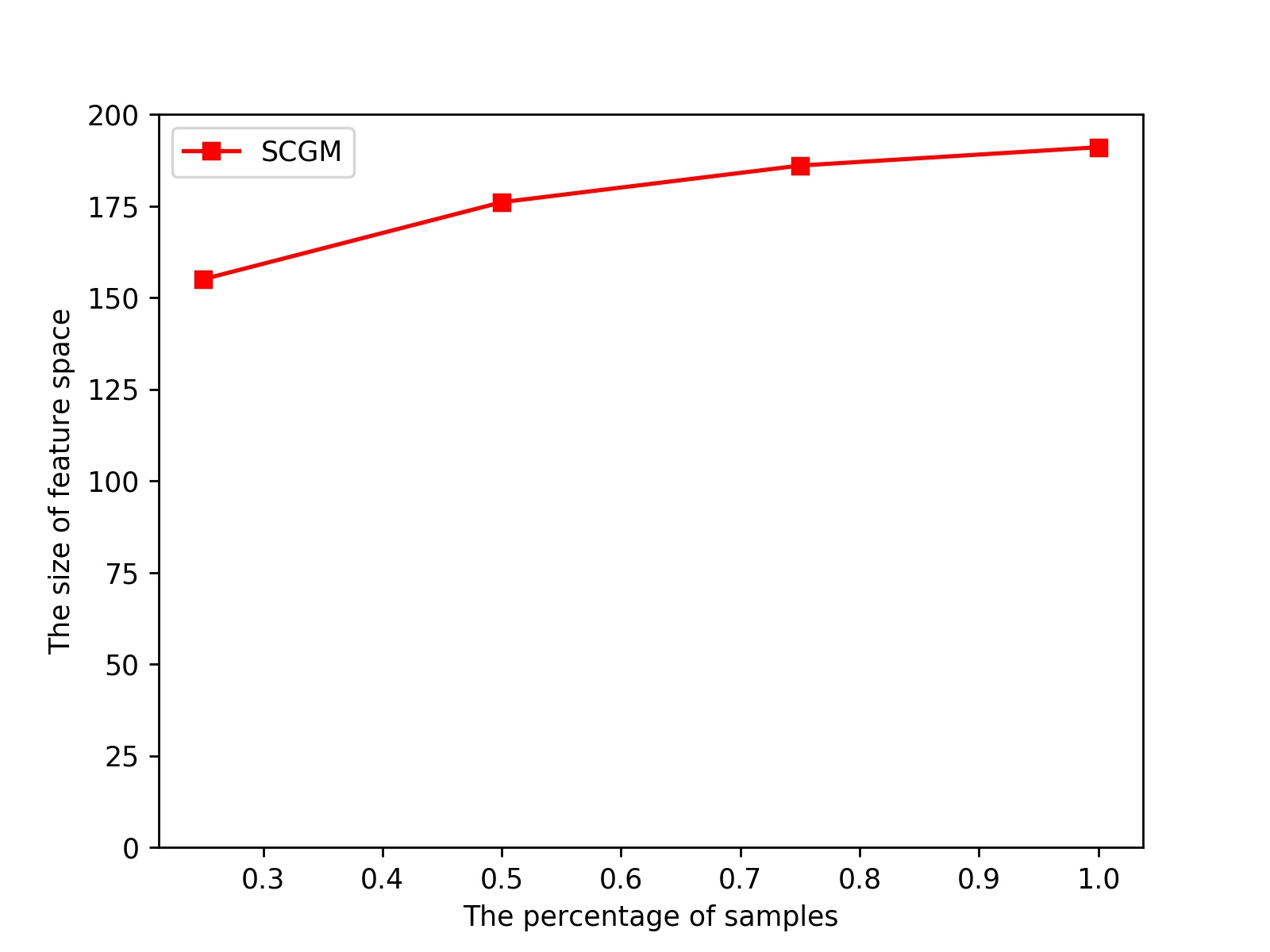}}
\caption{The analysis of feature space between 4-gram model and SCGM.}     
\label{fig}     
\end{figure}

Because the SCGM groups system calls based on reachability relation, it can have the less number of features compared to 4-gram model, which can relieve malware detection overheads. For instance, for the system call trace shown in Figure 2, the number of features generated by 4-gram model is 15, while the number of features generated by SCGM is 5, as listed in Table 6.
\begin{table}
\centering
\small
\centering
\caption{For the system call trace example shown in Figure 2, the features generated by 4-gram model and SCGM, repectively.}\label{tab1}
\begin{tabular}{|c|p{10cm}<{\centering}|}
\hline
Techniques &  Features\\
\hline
 &  $\left\{openat, brk, brk, getdents64\right\}$, $\left\{brk, brk, getdents64, open\right\}$, \\
&  $\left\{brk, getdents64, open, fstat64\right\}$, $\left\{getdents64, open, fstat64, read\right\}$,\\
&  $\left\{open, fstat64, read, open\right\}$, $\left\{fstat64, read, open, fstat64\right\}$,\\
4-gram & $\left\{read, open, fstat64, read\right\}$, $\left\{open, fstat64, read, mmap\right\}$,\\
 &  $\left\{fstat64, read, mmap, open\right\}$, $\left\{read, mmap, open, open\right\}$, \\
&  $\left\{mmap, open, open, read\right\}$, $\left\{open, open, read, write\right\}$,\\
&  $\left\{open, read, write, chmod\right\}$, $\left\{read, write, chmod, utime\right\}$, \\
&  $\left\{write, chmod, utime, execve\right\}$\\
\hline
SCGM & $\left\{openat, getdents64\right\}$, $\left\{brk\right\}$, $\left\{mmap\right\}$, $\left\{ open, fstat64, read\right\}$, \\
 & $\left\{open, open, read, write, chmod, utime, execve\right\}$\\
\hline
\end{tabular}
\end{table}

\subsubsection{Malware detection performance:}
Moreover, SCGDet can capture the semantic features of run-time program. For example, the feature, $\left\{open, open,\right.$ 
$\left.read, write, chmod, utime, execve\right\}$, can represent the typical propagation behavior of virus. While n-gram model only considers the strict order of system calls, which makes SCGDet have the higher detection accuracy and the lower false positives. The detailed experimental results are shown in Table 7. Because RF is an ensemble learning algorithm, it has the best malware detection performance. With RF as the classifier, SCGDet can have a TPR of 97.44\%, an FPR of 1.96\% and an Accuracy of 97.78\% in our dataset. In general, SCGDet can improve the average TPR of 2.86\%, reduce the average FPR of 14.75\% and improve the average Accuracy of 8.887\% as shown in Figure 6.
\begin{table}
\small
\centering
\caption{The comparison of experimental results between 4-gram technique and SCGDet technique.}
\begin{tabular}{|p{2cm}<{\centering}|c|c|c|c|c|c|c|c|}
\hline
Technique & \multicolumn{4}{|c|}{4-gram \cite{r6} \cite{r12}} & \multicolumn{4}{|c|}{SCGDet}\\
\hline
Classifier & LR(\%)	& SVM(\%) & RF(\%) & Average &	LR(\%)	& SVM(\%) & RF(\%) & Average\\
\hline
TPR	& 92.86	& 92.16	& 95.12	& 93.38	& 95.45	& 95.83	& 97.44	& 96.24\\
\hline
FPR	& 20.83	& 17.95	& 16.33	& 18.37	& 6.52	& 2.38	& 1.96	& 3.62\\
\hline
ACC	& 85.56	& 87.78	& 88.89	& 87.41	& 94.44	& 96.67	& 97.78	& 96.297\\
\hline
\end{tabular}
\end{table}

\begin{figure}
\centering
\includegraphics[width=10cm, height=6.5cm]{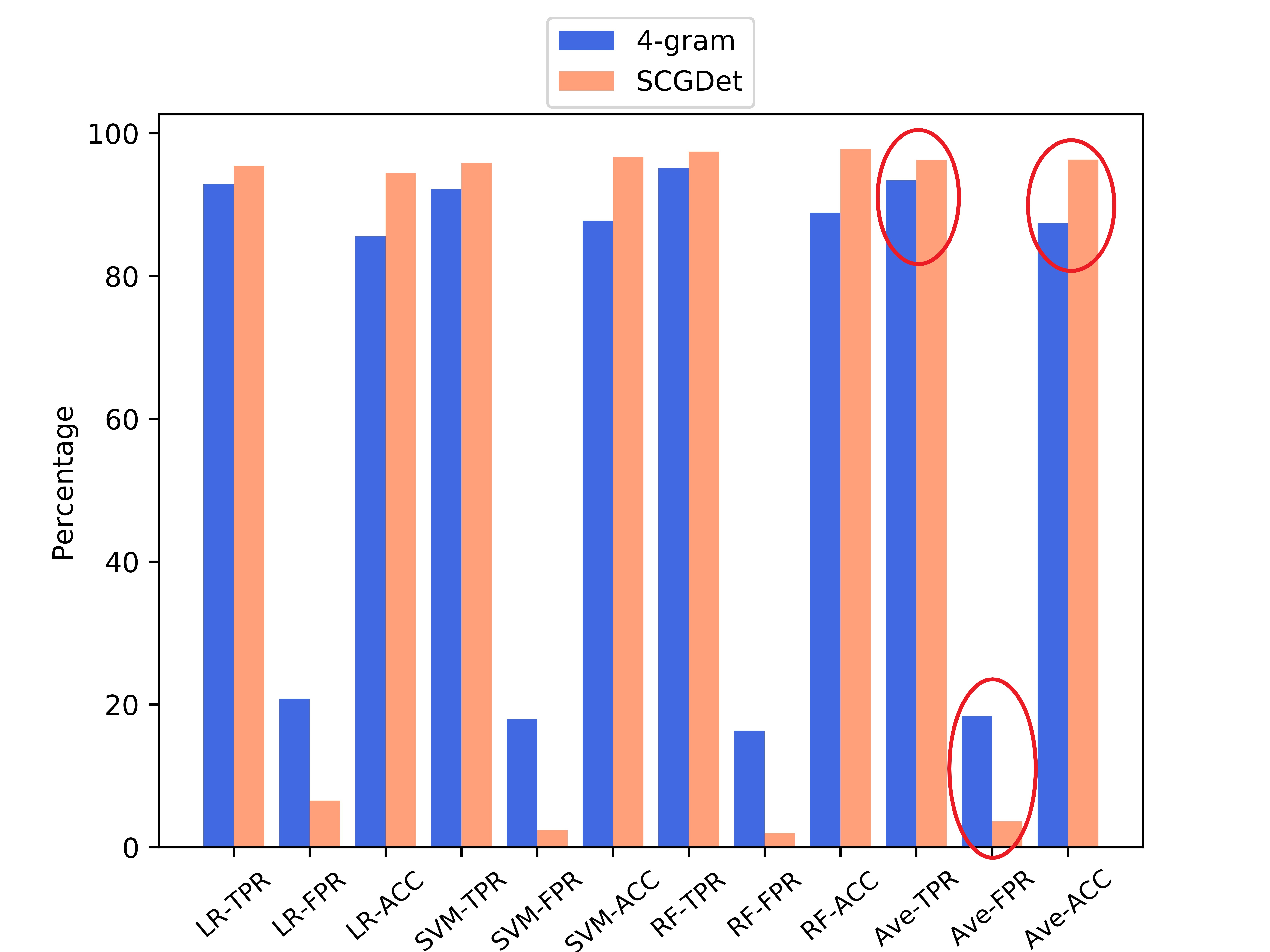}
\caption{The comparison of malware detection performances between 4-gram technique and SCGDet technique} \label{fig1}
\end{figure}

\section{Related works}
In the past, a considerable number about malware detection researches have been published. The security researchers have proposed many feature-statistics based approaches to identify malware. The statistics of features include the frequency, the prior probability, the entropy and information gain and so on. \cite{r13} proposed an integrated approach that use static and dynamic features to classify the benign and malware samples. They applied four common machine learning classifiers to carry out their objective. Ahmadi et al. \cite{r14} also proposed a similar approach, which use API calls to conduct their feature construction. In addition, they applied the feature selection methods to remove the redundant and irrelevant features. \cite{r15} proposed the visual analysis of malware behavior using treemaps and threaded graphs and used the temporal values of system object for malware detection. However, this method can lead to high false positive rates as proposed statistics have a low signal-to-noise ratio.

The n-gram is a popular data-mining technique, and has been applied in dynamic malware detection approaches. \cite{r16} presented an approach that can identify the malicious behaviors using Application Programming Interface (API) and system calls respectively. They applied a signature-like approach in order to match the n-gram features that are obviously present in malware but absent in benign programs. Wu and Yap \cite{r17} presented a visualization approach to cluster those malware samples that have similar malicious behaviors. They made use of byte opcodes and constructed the n-gram features, and they also applied hash-based technique to reduce the feature space. Because the syetem calls can provide effective information about the runtime behaviors of the program, many malware detection techniques based on system call have been proposed. Forrest et al. \cite{r18} first proposed anomaly detection using the sequences of system calls. Then \cite{r19} used system call arguments and sequences to capture interrelation among different arguments of the system call, which can detect the abnormal behaviors of run-time programs. Canali et al. \cite{r6} applied common data-mining technique, such as n-gram, m-bag and k-tuples, to conduct the feature construction and used machine learning methods to predict the program under detection. 

However, these n-gram based malware detection methods can lead to substantial performance overhead due to the huge feature space. Moreover, these methods do not capture the high-level semantic information of malicious behaviors and do not consider the frequency of system call patterns, which can result in high false positives. For example, malware can invoke repetitively those security-critical system call patterns to perform malicious attack such as denial of service and CPU starvation. In this paper, we propose the system call pruning method to filter out system calls that have little impact on malware detection, and propose the SCGM for feature construction. Our proposed approach, SCGDet, can capture the semantic features of typical malware behaviors by grouping system calls based on reachability relation and considering the frequency of system call patterns. Compared to traditional n-gram based method, SCGDet reduces significantly the feature space and has lower false positive rate.

\section{Conclusion}
In this paper, we first observed that malware usually invokes security-critical system calls to achieve malicious activities, and observed that malware and its variants belonging to the same family usually invoke similar system call patterns due to performing the similar malicious behaviors. Therefore, we first develop a system call pruning method to exclude system calls that have few impact on malware detection. Then we propose a novel feature construction method, system call graph model (SCGM), to capture the semantic features of typical malware behaviors by grouping system calls based on reachability relation and considering the frequency of system call patterns. Compare to the n-gram method, our propose technique, SCGDet, has the smaller feature space, the lower false positive rates and high detection accuracy. Experimental results show that SCGDet can reduce the average FPR of 14.75\% and improve the average Accuracy of 8.887\%. In the future, we are purposed to study the classification among different malware to determine which malicious family the malware sample under detection belongs to.

%
%
%

\begin{thebibliography}{10}
\providecommand{\url}[1]{\texttt{#1}}
\providecommand{\urlprefix}{URL }
\providecommand{\doi}[1]{https://doi.org/#1}

\bibitem{r1}
Mcafee labs threats report september 2018.
  \url{https://www.mcafee.com/enterprise/en-us/assets/reports/rp-quarterly-threats-sep-2018.pdf}

\bibitem{r10}
Virusshare. \url{https://virusshare.com/}

\bibitem{r11}
Virustotal. \url{https://www.virustotal.com/#/home}

\bibitem{r14}
Ahmadi, M., Sami, A., Rahimi, H., Yadegari, B.: Malware detection by
  behavioural sequential patterns. Computer Fraud \& Security
  \textbf{2013}(8),  11--19 (2013)

\bibitem{r4}
Bayer, U., Moser, A., Kruegel, C., Kirda, E.: Dynamic analysis of malicious
  code. Journal in Computer Virology  \textbf{2}(1),  67--77 (2006)

\bibitem{r6}
Canali, D., Lanzi, A., Balzarotti, D., Kruegel, C., Christodorescu, M., Kirda,
  E.: A quantitative study of accuracy in system call-based malware detection.
  In: Proceedings of the 2012 International Symposium on Software Testing and
  Analysis. pp. 122--132. ACM (2012)

\bibitem{r5}
Christodorescu, M., Jha, S., Kruegel, C.: Mining specifications of malicious
  behavior. In: Proceedings of the the 6th joint meeting of the European
  software engineering conference and the ACM SIGSOFT symposium on The
  foundations of software engineering. pp. 5--14. ACM (2007)

\bibitem{r7}
Egele, M., Kruegel, C., Kirda, E., Yin, H., Song, D.: Dynamic spyware analysis
  (2007)

\bibitem{r2}
Egele, M., Scholte, T., Kirda, E., Kruegel, C.: A survey on automated dynamic
  malware-analysis techniques and tools. ACM computing surveys (CSUR)
  \textbf{44}(2), ~6 (2012)

\bibitem{r18}
Forrest, S., Hofmeyr, S.A., Somayaji, A., Longstaff, T.A.: A sense of self for
  unix processes. In: Proceedings 1996 IEEE Symposium on Security and Privacy.
  pp. 120--128. IEEE (1996)

\bibitem{r13}
Islam, R., Tian, R., Batten, L.M., Versteeg, S.: Classification of malware
  based on integrated static and dynamic features. Journal of Network and
  Computer Applications  \textbf{36}(2),  646--656 (2013)

\bibitem{r8}
Kolbitsch, C., Comparetti, P.M., Kruegel, C., Kirda, E., Zhou, X.y., Wang, X.:
  Effective and efficient malware detection at the end host. In: USENIX
  security symposium. vol.~4, pp. 351--366 (2009)

\bibitem{r16}
Lanzi, A., Balzarotti, D., Kruegel, C., Christodorescu, M., Kirda, E.:
  Accessminer: using system-centric models for malware protection. In:
  Proceedings of the 17th ACM conference on Computer and communications
  security. pp. 399--412. ACM (2010)

\bibitem{r19}
Maggi, F., Matteucci, M., Zanero, S.: Detecting intrusions through system call
  sequence and argument analysis. IEEE Transactions on Dependable and Secure
  Computing  \textbf{7}(4),  381--395 (2010)

\bibitem{r12}
Mehdi, S.B., Tanwani, A.K., Farooq, M.: Imad: in-execution malware analysis and
  detection. In: Proceedings of the 11th Annual conference on Genetic and
  evolutionary computation. pp. 1553--1560. ACM (2009)

\bibitem{r9}
Pedregosa, F., Varoquaux, G., Gramfort, A., Michel, V., Thirion, B., Grisel,
  O., Blondel, M., Prettenhofer, P., Weiss, R., Dubourg, V., Vanderplas, J.,
  Passos, A., Cournapeau, D., Brucher, M., Perrot, M., Duchesnay, E.:
  Scikit-learn: Machine learning in {P}ython. Journal of Machine Learning
  Research  \textbf{12},  2825--2830 (2011)

\bibitem{r15}
Trinius, P., Holz, T., G{\"o}bel, J., Freiling, F.C.: Visual analysis of
  malware behavior using treemaps and thread graphs. In: 2009 6th International
  Workshop on Visualization for Cyber Security. pp. 33--38. IEEE (2009)

\bibitem{r17}
Wu, Y., Yap, R.H.: Experiments with malware visualization. In: International
  Conference on Detection of Intrusions and Malware, and Vulnerability
  Assessment. pp. 123--133. Springer (2012)

\bibitem{r3}
You, I., Yim, K.: Malware obfuscation techniques: A brief survey. In: 2010
  International conference on broadband, wireless computing, communication and
  applications. pp. 297--300. IEEE (2010)

\end{thebibliography}
%

\bibliographystyle{splncs04}

\end{document}